\documentclass[prd,aps,amsfonts,eqsecnum,superscriptaddress,nofootinbib,notitlepage,longbibliography,final]{revtex4-1}

\usepackage{amsmath, amssymb, amsthm}
\usepackage[margin=1in]{geometry}

\usepackage{graphicx}
\usepackage{xcolor}
\usepackage{tikz}
\usepackage{quantikz}

\usepackage[colorlinks=true, allcolors=blue]{hyperref}
\usepackage{cleveref}

\usepackage{ulem}
\usepackage{subcaption}
\usepackage{float}

\usepackage{soul}
\usepackage{cancel}

\newcommand{\dd}{\mathrm{d}}

\newtheorem{theorem}{Theorem}[section]

\newtheorem{lemma}[theorem]{Lemma}

\theoremstyle{definition}

\newtheorem{remark}[theorem]{Remark}

\usepackage{algpseudocode}
\usepackage{mdframed}

\begin{document}

\title{Preconditioned Multivariate Quantum Solution Extraction}

\author{Gumaro Rend\'on}
\email{grendon@fujitsu.com}
\affiliation{Fujitsu Research of America, Inc, Santa Clara, CA 95054, USA}

\author{\v{S}t\v{e}p\'an \v{S}m\'\i d}
% \thanks{\texttt{s.smid23@imperial.ac.uk}}}
\affiliation{Department of Computing, Imperial College London, United Kingdom}
\affiliation{Fujitsu Research of Europe Ltd., Slough SL1 2BE, United Kingdom}

\begin{abstract}
    Numerically solving partial differential equations is a ubiquitous computational task with broad applications in many fields of science. Quantum computers can potentially provide high-degree polynomial speed-ups for solving PDEs, however many algorithms simply end with preparing the quantum state encoding the solution in its amplitudes. Trying to access explicit properties of the solution naively with quantum amplitude estimation can subsequently diminish the potential speed-up. In this work, we present a technique for extracting a smooth positive function encoded in the amplitudes of a quantum state, which achieves the Heisenberg limit scaling. We improve upon previous methods by allowing higher dimensional functions, by significantly reducing the quantum complexity with respect to the number of qubits encoding the function, and by removing the dependency on the minimum of the function using preconditioning. Our technique works by sampling the cumulative distribution of the given function, fitting it with Chebyshev polynomials, and subsequently extracting a representation of the whole encoded function. Finally, we trial our method by carrying out small scale numerical simulations.
\end{abstract}

\maketitle

%%%%%%%%%%%%%%%%%%%%%%%%%%%%%%%%%%%%%%%%%%%%%%%%%%%%%%%%%%%%%%%%%%%%%%%%%%%%%%%%

\section{Introduction}

There is a well-known and often understated challenge in quantum algorithms for linear systems and differential equations: \textbf{solution extraction}. This issue traces back to the landmark work of Harrow, Hassidim, and Lloyd~\cite{harrow2009quantum}, where the authors achieved an exponential speedup for matrix inversion over the best known classical algorithms, with respect to system size. The key assumption enabling this exponential improvement was that we do not need to recover the entire solution vector explicitly, but rather we only require global properties—such as expectation values of observables—that can be measured efficiently on a quantum computer. Indeed, they argue that attempting to access every component of the solution vector would nullify the exponential advantage, as it would require a number of queries scaling exponentially with the number of qubits.

This paradigm has been followed by many subsequent quantum linear systems solvers and quantum algorithms for differential equations (see, e.g., \cite{Berry_2014,childs2017quantum,berry2017quantum,Childs2021highprecision}). In these works, the solution is typically prepared as a quantum state but not fully read out; instead, it is assumed that the relevant quantities can be extracted efficiently. However, this assumption sidesteps a crucial difficulty: if one desires \textbf{explicit access} to the solution, even at a single coordinate, the corresponding probability amplitude is typically suppressed by a factor of $1/2^n$, where $n$ is the number of qubits representing the discretized domain.

There is a first attempt at solving this problem in the work \cite{rendon2025_SE}.
The author considers an unknown normalized function $\psi(x)$ over a compact interval:
\begin{align}
    \int_{-1}^{1} |\psi(x)|^2 \, {\rm d}x = 1,
\end{align}
with the following bounds on its derivatives:
\begin{align}
    \bigg|\frac{{\rm d}^k \psi(x)}{{\rm d}x^k}\bigg| \leq \Lambda^{k+1}.
\end{align}
It is supposed that one can encode $\psi(x)$ as a quantum state using $n$ qubits:
\begin{align}
    \langle 0 |_a U_{\psi} | 0 \rangle | 0 \rangle_a = a_\psi |\psi \rangle,
\end{align}
where
\begin{align}
    |\psi\rangle = \frac{1}{\sqrt{\mathcal{N}}} \sum_{j=0}^{2^n-1} \psi(x_j) |j\rangle,
\end{align}
with normalization factor $\mathcal{N} \approx 2^{n-1}$ and grid points $x_j = 2 j/2^n - 1$. The ancilla register $\ket{0}_a$ and amplitude $a_\psi$ encode the success probability of state preparation.

In this representation, measuring a single computational basis state $|j\rangle$ returns the corresponding probability weight $|\psi(x_j)|^2 / 2^n$. This exponential suppression implies that naïve sampling requires $O(2^n)$ repetitions to estimate a single value with constant precision. Even with quantum amplitude estimation, which yields a quadratic improvement~\cite{brassard2000}, the cost still scales like $O\!\left(|\psi(x_j)|/\sqrt{2^n}\right)$. Classical shadow tomography and other state-learning techniques~\cite{huang2020predicting} do not circumvent this issue, as their sample complexity depends polynomially on $1/\epsilon$, where $\epsilon$ is the additive precision, leaving the $2^{-n}$ factor unmitigated.

There are other works using tensor networks and interpolating polynomials (See \cite{Miyamoto:2022wrw}) to extract the function encoded in a state vector. The polynomial degree required has a good scaling, however, the method to extract this polynomial relies on an optimization problem where the correct convergence is not guaranteed (or one could guarantee it, but the cost grows exponentially with the number of degrees of freedom).

Here, we follow the work in ~\cite{rendon2025_SE} where we have guarantees of finding the approximating polynomial for a target precision. 

What is achieved in \cite{rendon2025_SE} is summarized in the following lemma:

\begin{lemma}\label{thm:SE_1D}
Provided an unknown analytic function function $\psi(x)$ for $x \in [-1,1] $ that is normalized:
\begin{align*}
    \int^{1}_{-1} \left|\psi(x)\right|^2 {\rm d} x = 1,
\end{align*}
and whose derivatives are bounded by $\left| \frac{{\rm d}^j\psi(x)}{d x^j} \right| \leq \Lambda^{j+1} $, which can be stored in a quantum memory with $n$ qubits the following way:
\begin{align*}
    \langle 0 |_a U_{\psi}|0 \rangle | 0 \rangle_a = a_{\psi} | \psi \rangle,
\end{align*}
where
\begin{align*}
    |\psi \rangle &= \frac{1}{\sqrt{\mathcal{N}}} \sum^{2^n-1}_{j=0} \psi(x_j) |j \rangle,
\end{align*}
 $\mathcal{N} = \sum_j |\psi(x_j)|^2 $, and $x_j = 2 j/2^n -1$, one can estimate it at a quantum gate cost that goes like
 \begin{align*}
    \tilde{O}\left(\frac{1}{a_{\psi}}\frac{\Lambda^3 n^4 }{\epsilon_{\rm total}} \frac{\max_{x} \psi(x) }{\min_x \psi(x)} \right).
\end{align*}
    
\end{lemma}

In this work, we revisit this long-standing bottleneck and develop methods to extract pointwise information about $\psi(\vec{x})$, the $D$-dimensional version of this problem. We improve the dependence of the method with respect to the condition number $\kappa = \frac{\max \psi(\vec{x})}{\min \psi (\vec{x})}$, by removing its dependence on it entirely. We do this by using a pre-conditioning method based on interference with a constant distribution (See Section \ref{sec:precond}). Moreover, we improve the extraction of a single integral point by using indicator function oracles (comparator circuits) ~\cite{wang2025comprehensive,cuccaro2004newquantumripplecarryaddition,yuan2023improvedqftbasedquantumcomparator} instead of the binary decomposition used in this previous work. The naive extension would have given a scaling of $n^D$ for a single point extraction, while the improved extraction gives ${\rm poly}(n,D)$ for a single point. The new algorithm is summarized in \Cref{fig:pseudo_alg}.

\begin{figure}
\caption{Pseudo-code for the Solution Extraction Algorithm}\label{fig:pseudo_alg}
\begin{mdframed}[linewidth=1pt, roundcorner=4pt]
\begin{algorithmic}[1]

\Require $U_{\psi}$ which block-encodes the solution $\langle 0 |_a U_{\psi} | 0 \rangle | 0 \rangle_a = a_\psi |\psi \rangle$
\Ensure Output description

\State If not known precisely from theory, determine $a_{\psi}$ through Quantum Amplitude Estimation (QAE)
\State Construct $U_{\tilde{\Psi}}$ which lifts the solution enough to have a condition number of order $1$
\State Construct the Grover operator to do quantum amplitude estimation to estimate the parameter $a_{\rm shift}$

\For{$j_1 \gets 1$ to $M$}
    \State $\ddots$
    \For{$j_D \gets 1$ to $M$}
        \State Construct Grover operator of the indicator function and $U_{\psi}$
        \State Estimate $\tilde{\Psi}(x_1,x_2,\dots,x_D)$ at $x_{\vec{j}}$ with said Grover operator at a target precision $\mathcal{O}\left(\epsilon/M^{3D}\right)$
    \EndFor
    \State With the sampled $\tilde{\Psi}$ at the $D$-dimensional Chebuyshev nodes, perform the Chebyshev interpolation (ie solve $a=V^{-1} f$, where $f$ are the samples of $\tilde{\Psi}$)
    \State Obtain $\tilde{\psi}^2$ from the derivative of the interpolant above
    \State Obtain the approximation of $\tilde{\psi}$ by taking the square root of the approximant of $\tilde{\psi}^2$ (Return an efficient way to evaluate the Taylor series for $\sqrt{\cdot}$, for example, or instead evaluate at each $\mathbf{x}$)
    \State Obtain the approximation $\psi$ from $\tilde{\psi}$ and the approximation of $a_{\rm shift}$ 
\EndFor

\State \Return Resulting function

\end{algorithmic}
\end{mdframed}
\end{figure}

%%%%%%%%%%%%%%%%%%%%%%%%%%%%%%%%%%%%%%%%%%%%%%%%%%%%%%%%%%%%%%%%%%%%%%%%%%%%%%%%

\section{Preconditioning}\label{sec:precond}

\subsection{Shift by a constant}

Consider a function $\psi:[-1,1]\to \mathbb{R}_+$ which is non-negative and normalized such that $\int_{-1}^1 \psi(x)^2\ \dd x = 1$. This function shall be encoded in the amplitudes of a quantum state. We wish to extract this function, but the complexity of this will depend on its condition number $\kappa = \frac{\max \psi(x)}{\min \psi(x)}$ \cite{rendon2025_SE}, and so we want to first decrease this condition number to prevent a potential exponential overhead in the case that $\min \psi(x)$ would approach $0$.

Assume we have access to a unitary gate $U_\psi$ which implements the state \begin{align}
    |\psi\rangle = U_\psi |\mathbf{0}\rangle = \frac{1}{\sqrt{\mathcal{N}}} \sum\limits_{j=0}^{2^n-1}\psi(x_j)|j\rangle\,,
\end{align} and also its controlled version $C\text{-}U_\psi$. Denote the uniform superposition by $|\phi\rangle = \frac{1}{\sqrt{2^n}} \sum\limits_{j=0}^{2^n-1}|j\rangle$. Then we can construct \begin{align}
    &(\alpha |0\rangle + \beta |1\rangle ) |\mathbf{0}\rangle \overset{\overline{C}-U_\psi}{\longrightarrow} \alpha |0\rangle|\psi\rangle  + \beta |1\rangle |\mathbf{0}\rangle \overset{C-U_\phi}{\longrightarrow} \alpha |0\rangle|\psi\rangle  + \beta |1\rangle |\phi\rangle \cr 
    &\quad \overset{H}{\longrightarrow} \frac{1}{\sqrt{2}} (|0\rangle (\alpha |\psi\rangle + \beta |\phi\rangle) + |1\rangle (\alpha |\psi\rangle - \beta |\phi\rangle))
\end{align} Assuming $\alpha,\beta$ to be real and positive (so $\beta = \sqrt{1-\alpha^2}$), we have that the probability of measuring $0$ on the first qubit is \begin{align}
    \mathbb{P}(0) = \frac{1}{2} + \alpha\beta \frac{1}{\sqrt{2^n \mathcal{N}}} \sum\limits_{j} \psi(x_j) > \frac{1}{2}\,.
\end{align} This will result in the post-measurement state \begin{align}
    |\chi\rangle &= \frac{1}{\sqrt{\mathcal{N}_2}} (\alpha |\psi\rangle + \beta |\phi\rangle ) 
    = \frac{\alpha}{\sqrt{\mathcal{N}\mathcal{N}_2}} \sum\limits_{j} \left( \psi(x_j) + \sqrt{\alpha^{-2} - 1} \cdot \sqrt{\frac{\mathcal{N}}{2^n}} \right)|j\rangle \\ 
    &\approx \frac{\alpha}{\sqrt{\mathcal{N}\mathcal{N}_2}} \sum\limits_{j} \left( \psi(x_j) + \frac{\sqrt{\alpha^{-2} - 1}}{\sqrt{2}} \right) |j\rangle
    \overset{\text{set}}{=} \frac{1}{\sqrt{\mathcal{N}_3}} \sum_j \widetilde{\psi}(x_j) |j\rangle\,,
\end{align} where $\widetilde{\psi}(x) = a_{\rm shift} \cdot \left( \psi(x) + \frac{\sqrt{\alpha^{-2} - 1}}{\sqrt{2}}\right)$ with $a$ such that $\int_{-1}^1 \widetilde{\psi}(x)^2\ \dd x =1$. The approximation here follows from $\mathcal{N}/2^{n-1} = 1 + \mathcal{O}\left(\frac{1}{2^n}\right)$. The value of the normalization is then \begin{align} a_{\rm shift} = \left(\alpha^{-2} + \sqrt{2}\sqrt{\alpha^{-2} - 1} \int_{-1}^1\psi(x)\ \dd x  \right)^{-1/2}\ \in [0,1]\,. \end{align}

Since we need to know the value of $a$ to recover $\psi(x)$ from $\widetilde{\psi}(x)$, we have to understand the value of the $\ell_1$ norm of $\psi$, i.e. the integral $\int_{-1}^1 \psi(x)\ \dd x$. 

\subsection{Estimation of $a_{\rm shift}$}

We define $c$ as the probability of measuring zero on the interference flag qubit:
\begin{align}
    c = \mathbb{P}(0) = \frac{1}{2} + \alpha\beta \frac{1}{\sqrt{2^n \mathcal{N}}} \sum\limits_{j} \psi(x_j)  > \frac{1}{2}\,.
\end{align} 

\begin{align}
    |\chi\rangle &= \frac{1}{\sqrt{\mathcal{N}_2}} (\alpha |\psi\rangle + \beta |\phi\rangle ) 
    = \frac{\alpha}{\sqrt{\mathcal{N}\mathcal{N}_2}} \sum\limits_{j} \left( \psi(x_j) + \sqrt{\alpha^{-2} - 1} \cdot \sqrt{\frac{\mathcal{N}}{2^n}} \right)|j\rangle \\ 
    &\approx \frac{\alpha}{\sqrt{\mathcal{N}\mathcal{N}_2}} \sum\limits_{j} \left( \psi(x_j) + \frac{\sqrt{\alpha^{-2} - 1}}{\sqrt{2}} \right) |j\rangle
    \overset{\text{set}}{=} \frac{1}{\sqrt{\mathcal{N}_3}} \sum_j \widetilde{\psi}(x_j) |j\rangle\,,
\end{align}

Thus,
\begin{align}
    \mathcal{N}_ 2 = 2 c.
\end{align}
We also have by definition:
\begin{align}
    \frac{\alpha}{ \sqrt{\mathcal{N} \mathcal{N}_2}} = \frac{a_{\rm shift}}{\sqrt{\mathcal{N}_3}}.
\end{align}
Now, because $\widetilde{\psi}(x)$ is also as smooth function, we know that $\mathcal{N}_3/2^{n-1} = 1 + \mathcal{O}\left(\frac{1}{2^n}\right)$. Thus,
\begin{align}
    a_{\rm shift} \approx \frac{\alpha}{\sqrt{2c}} = \mathcal{O}(1).
\end{align}
Thus, we need only determine $\sqrt{c}$ with an error $\mathcal{O}(\epsilon)$ to obtain $a$ with an error that is also $\mathcal{O}(\epsilon)$. For the case of unitary encoding of the function $\psi(x)$, the cost of this would scale like $\mathcal{O}(1/\epsilon)$.

\subsection{Bounds on the condition number $\kappa(\widetilde{\psi})$}

Observe that, if we further assume $\left| \frac{\dd \psi(x)}{\dd x} \right| \leq \Lambda^2$, we can bound the $\ell_1$ norm by \begin{align}
    \sqrt{2-\frac{4}{3}\Lambda^4} \leq \int\limits_{-1}^1 \psi(x)\ \dd x \leq \sqrt{2}\,.
\end{align}
To get the upper bound, simply observe $(\psi(x)-\psi(y))^2 \geq 0$. Hence get \begin{align}
    0 \leq \int\limits_{-1}^1 \int\limits_{-1}^1 (\psi(x)-\psi(y))^2\ \dd x\dd y = 4  \int\limits_{-1}^1\psi(x)^2\ \dd x - 2 \left( \int\limits_{-1}^1 \psi(x)\ \dd x \right)^2 = 4  - 2 \left( \int\limits_{-1}^1 \psi(x)\ \dd x \right)^2\,.
\end{align} 
For the lower bound, since $\psi(x)$ is assumed differentiable, Mean Value Theorem tells us that $\psi(x)-\psi(y) = (x-y) \cdot \psi'(z)$ for some $z \in [x,y]$. Hence we get that $(\psi(x)-\psi(y))^2 \leq (x-y)^2 (\max_z \psi'(z))^2 \leq (x-y)^2 \Lambda^4$. Integrating this expression gives \begin{align}
    \int\limits_{-1}^1 \int\limits_{-1}^1 (\psi(x)-\psi(y))^2\ \dd x\dd y = 4   - 2 \left( \int\limits_{-1}^1 \psi(x)\ \dd x \right)^2 \leq \frac{8}{3} \Lambda^4\,.
\end{align}
Note that this bound is useful only for $\Lambda \leq \left(\frac{3}{2}\right)^{1/4} \approx 1.107$, otherwise it becomes vacuously true as $\int\psi\ \dd x \geq 0$.
Hence we find that \begin{align}
     \left(\alpha^{-2} + 2\sqrt{\alpha^{-2} - 1}   \right)^{-1/2} \leq a \leq  \left(\alpha^{-2} + \sqrt{\alpha^{-2} - 1} \sqrt{4-\frac{8}{3}\Lambda^4}  \right)^{-1/2} \leq  \alpha\,.
\end{align}
Taking $\tilde a$ to be the midpoint of these two bounds, $\tilde{a} = \left(\alpha + \left(\alpha^{-2} + 2\sqrt{\alpha^{-2} - 1}   \right)^{-1/2}\right)/2$, we have that $|a-\tilde a| \leq 0.12$ and $|\tilde a /a - 1| \leq 0.21$, hence incurring only a fixed relative error when extracting $\psi(x)$ using this value instead of $a$.

The new condition number after this shift is then \begin{align}
    \kappa = \frac{\max \widetilde{\psi}(x)}{\min \widetilde{\psi}(x)} = \frac{(\max \psi(x)) + \frac{\sqrt{\alpha^{-2} - 1}}{\sqrt{2}}}{(\min \psi(x)) + \frac{\sqrt{\alpha^{-2} - 1}}{\sqrt{2}}} \leq \frac{\max \psi(x)}{\min \psi(x)}\,,
\end{align} which is by construction an $\mathcal{O}(1)$ quantity as $\min\widetilde{\psi}(x)$ is bounded away from $0$.

Note that since we are shifting all amplitudes uniformly, this preconditioning step directly generalizes to the case of a $D$-dimensional function $\psi$ as encoded in \eqref{eqn:D-dim function}.

%%%%%%%%%%%%%%%%%%%%%%%%%%%%%%%%%%%%%%%%%%%%%%%%%%%%%%%%%%%%%%%%%%%%%%%%%%%%%%%%

\section{Multivariate solution extraction}

Let $\psi:[-1,1]^D\to \mathbb{R}_+$, and assume oracle access for preparing the quantum state $|\psi\rangle = U_\psi |\mathbf{0}\rangle$ given by \begin{align}\label{eqn:D-dim function}
    |\psi\rangle = \frac{1}{\sqrt{\mathcal{N}}} \sum_{j_1,\dots,j_D = 0}^{2^n-1} \psi\left(x^{(j_1)},\dots,x^{(j_D)}\right) |j_1\rangle \dots |j_D\rangle\,,
\end{align} where $x^{(j)} = \frac{j}{2^{n-1}}-1$. Assume the function to be normalized like $\int_{[-1,1]^D} \psi(\mathbf{x})^2\ \dd x^D = 1$. If we were to use quantum amplitude estimation on the state $\ket{\psi}$ directly, we would run into issues with the encoded functional values being exponentially subnormalized, and hence exponentially difficult to extract with a constant error.

\subsection{Integral estimation}

Define the square integral of $\psi$ by \begin{align}
    \Psi(x_1,\dots,x_D) = \int_{-1}^{x_1} \dots \int_{-1}^{x_D} \psi(y_1,\dots,y_D)^2\ \dd y^D\,.
\end{align}
Estimating this integral instead of the amplitudes avoids the subnormalization issues since the integral represents their cumulative distribution. We can also sample this integral directly via QAE with an indicator function defined by \begin{align}
    f_\mathbf{k}(\mathbf{j}) = \mathbf{1}(j_1 \leq k_1-1) \dots \mathbf{1}(j_D \leq k_D-1)\,.
\end{align} In other words, we want to use quantum phase estimation with the input state $\ket{\psi}$ on the Grover amplitude amplification operator, where $\mathbf{f_k}$ indicates the good subspace. We can implement the quantum oracle $U_{f_\mathbf{k}}$ using $\mathcal{O}(Dn)$ Toffoli gates in $\mathcal{O}(Dn)$ qubits. The basic implementation of this so-called quantum comparator is discussed e.g.~in \cite{cuccaro2004newquantumripplecarryaddition}, with a different variant using a single ancilla (but a deeper circuit) in \cite{yuan2023improvedqftbasedquantumcomparator}. For our $D$-dimensional comparison, we just need to implement one for each of the $D$ registers and then take $C^D\text{-}NOT$ controlled by all the outputs, followed by uncomputing the intermediate results. Running quantum amplitude estimation on the state $|\psi\rangle$ with this indicator function yields an estimate to \begin{align} p 
    &= \frac{1}{\mathcal{N}} \sum_{j_1=0}^{k_1-1}\dots \sum_{j_D=0}^{k_D-1} \psi \left(x^{(j_1)},\dots,x^{(j_D)}\right)^2\,.%\\
    %&=\Psi(x^{(k_1)},\dots,x^{(k_D)})+ \mathcal{O}\left(D \cdot \frac{\Lambda^2}{2^n}\right)\,.
\end{align} 
Note that we can approximate the integral $\Psi$ using a Riemann sum:
\begin{align}
    \Psi\left(x^{(k_1)},\dots,x^{(k_D)}\right) &= \int_{-1}^{x^{(k_1)}} \dots \int_{-1}^{x^{(k_D)}} \psi(y_1,\dots,y_D)^2\ \dd y^D\\
    &= \frac{1}{2^{D\cdot(n-1)}} \sum_{j_1=0}^{k_1-1}\dots \sum_{j_D=0}^{k_D-1} \psi \left(x^{(j_1)},\dots,x^{(j_D)}\right)^2 + \mathcal{O}\left(D \cdot \frac{\Lambda^2}{2^n}\right)
\end{align}
From the normalization $\Psi(1,\dots,1)=1$ it also follows that \begin{align}
    \frac{\mathcal{N}}{2^{D\cdot(n-1)}} = \frac{1}{2^{D\cdot(n-1)}}\sum_{j_1,\dots,j_D = 0 }^{2^n-1} \psi\left(x^{(j_1)}, \dots, x^{(j_D)}\right)^2 = 1 + \mathcal{O}\left(D \cdot \frac{\Lambda^2}{2^n}\right)\,,
\end{align} hence we can approximate the integral $\Psi$ by $p$ since \begin{align}
    \Psi\left(x^{(k_1)},\dots,x^{(k_D)}\right)
    = p + \mathcal{O}\left(D \cdot \frac{\Lambda^2}{2^n}\right)\,.
\end{align}

We wish to use quantum phase estimation on the Grover rotation operator \begin{align}
    W = - U R_{|0\rangle} U^\dagger R_{f_\mathbf{k}}\,,
\label{eqn:Grover operator}\end{align} where the reflection $R_{f_\mathbf{k}}$ is implementable with one call of $U_{f_\mathbf{k}}$. The gate complexity of $W$ is hence $\mathcal{O}(Dn+ \operatorname{complexity}(U))$. 

Now to estimate $p$ within a fixed accuracy $\epsilon$ and success probability $8/\pi^2$, we need $T$ calls of the operator $W$, where $T$ obeys \begin{align}
    \epsilon \leq \frac{2\pi\sqrt{p(1-p)}}{T} + \frac{\pi^2}{T^2}\,.
\end{align} This is maximized at $p=0.5$, and hence taking \begin{align}
    T \geq \pi \frac{1+\sqrt{1+4\epsilon}}{\epsilon} = \mathcal{O}\left(\frac{1}{\epsilon}\right)
\end{align} ensures error on $p$  of at most $\epsilon$.

Overall, we find that we can sample $\Psi(x_1,\dots,x_D)$ within an error $\epsilon$ using $\mathcal{O}\left(\frac{Dn+\operatorname{complexity}(U)}{\epsilon}\right)$ gate complexity (assuming access to $U$, $U^\dagger$, and $C\text{-}U$).

\subsection{Chebyshev interpolation of the integral}

Now we shall consider interpolating a multivariate function with Chebyshev polynomials. Since we are working on a hypercubic domain $\mathcal{D} = [-1,1]^D$, the multivariate Chebyshev polynomials will be obtained in a tensorial fashion from the univariate, writing \begin{align}\label{eq:interp} \Psi(x_1,\dots,x_D) \approx P_{M-1} \Psi(x_1,\dots,x_D) = \sum_{j_1,\dots, j_D = 0}^{M-1} a_{j_1\dots j_D} \cdot t_{j_1}(x_1) \dots t_{j_D}(x_D)\,,
\end{align} where \begin{align}
t_j(x) = \begin{cases}
    \sqrt{1/M} \cdot T_0(x) & \text{if } j =0,\\
    \sqrt{2/M} \cdot T_j(x) & \text{if } j \in \{1,\dots,M-1\}
\end{cases}
\end{align} are the normalized Chebyshev polynomials. We will consider interpolation over Chebyshev nodes, given by \begin{align}
    \mathbf{x}_{\textup{cheb},\mathbf{k}} = \left(\cos\left(\frac{2k_1-1}{2M}\pi\right), \dots ,\cos\left(\frac{2k_D-1}{2M}\pi\right)\right)\,, \qquad k_i \in \{1,2,\dots,M\}\,.
\end{align}

First, we note from the proof of \cite[Theorem 11 p. 95]{bochner1948several} that the coefficients of the multivariate Chebyshev expansion decay like
\begin{align}
    \left|\tilde{a}_{j_1\dots j_D} \right| = \mathcal{O} \left(F \rho^{-j_1} \rho^{-j_2}\dots \rho ^{-j_D}\right)\,,
\end{align}
where $\max_{E(\mathbf{\rho})}|f(z)| \leq F$. The $\tilde{(\cdot)}$ on the coefficients denotes that they are normalized such that $f(x)=\sum_{j_1\dots j_D} \tilde{a}_{j_1\dots j_D} T_{j_1}(x_1) \dots T_{j_D}(x_D)$ with $\int^{1}_{-1} T_{i}(x) T_{j} (x) \frac{1}{\sqrt{1-x^2}} {d} x = \delta_{ij}$.

We also have that
\begin{align}
\| \partial^{\alpha} \psi \|_{\infty} \leq \Lambda^{|\alpha|+1},
\end{align}
Thus, we can bound $\psi$ on the circular poly-cylinders $D(\mathbf{r})$ containing the elliptical poly-cylinders $E(\mathbf{\rho})$ through
\[
\max_{z \in E(\mathbf{\rho})}
|\psi(z)| \leq \max_{z \in D(\mathbf{r})}
|\psi(z)| \;\le\; \Lambda\, e^{d\Lambda r},
\]
where $\rho = r \pm \sqrt{r^2 -1}$. Thus, $\rho \geq r$. The proof is as follows:

\begin{proof}

Let $\psi:\mathbb{R}^d \to \mathbb{C}$ extend analytically to $\mathbb{C}^d$, and assume
\[
\|\partial^\alpha \psi\|_{\infty} \le \Lambda^{|\alpha|+1}
\qquad\text{for all multi-indices }\alpha \in \mathbb{N}^d.
\]
Fix $x_0 \in \mathbb{R}^d$ and consider the polydisk/polycylinder:
\[
P_r(x_0)
=
\bigl\{ z \in \mathbb{C}^d : |z_j - x_{0,j}| \le r,\; j = 1,\dots,d \bigr\}.
\]
The multivariate Taylor expansion of $\psi$ around $x_0$ is
\[
\psi(z)
=
\sum_{\alpha\in\mathbb{N}^d}
\frac{\partial^\alpha \psi(x_0)}{\alpha!}\,(z-x_0)^\alpha,
\]
where $\alpha! = \alpha_1! \cdots \alpha_d!$ and
$(z-x_0)^\alpha = \prod_{j=1}^d (z_j - x_{0,j})^{\alpha_j}$.

For $z\in P_r(x_0)$ one has $|(z-x_0)^\alpha|\le r^{|\alpha|}$, hence
\[
|\psi(z)|
\le
\sum_{\alpha}
\frac{|\partial^\alpha\psi(x_0)|}{\alpha!}\, r^{|\alpha|}
\le
\sum_{\alpha}
\frac{\Lambda^{|\alpha|+1}}{\alpha!}\, r^{|\alpha|}
=
\Lambda
\sum_{\alpha}
\frac{(\Lambda r)^{|\alpha|}}{\alpha!}.
\]

Grouping by the total order $n = |\alpha|$,
\[
|\psi(z)|
\le
\Lambda \sum_{n=0}^{\infty} (\Lambda r)^n
\sum_{|\alpha|=n} \frac{1}{\alpha!}.
\]
Using the multinomial theorem:
\[
(1+\cdots+1)^n = d^n
= \sum_{|\alpha|=n} \frac{n!}{\alpha!},
\qquad\Rightarrow\qquad
\sum_{|\alpha|=n} \frac{1}{\alpha!}
= \frac{d^n}{n!},
\]
we obtain
\[
|\psi(z)|
\le
\Lambda \sum_{n=0}^{\infty} \frac{(d\Lambda r)^n}{n!}
=
\Lambda\, e^{d\Lambda r}.
\]
Thus, for every $z \in P_r(x_0)$,
\[
|\psi(z)| \;\le\; \Lambda\, e^{d\Lambda r}.
\]

\end{proof}

We want to bound the truncation error from the Chebyshev series of $\psi^2(x)$ of order $M$ on each variable. We obtain this series by first interpolating $\Psi(x)$ and then differentiating the interpolant. The error on $\psi^2(x)$ from truncating the Chebyshev series of $\Psi(x)$ is bounded through
\begin{align}
    \epsilon_{\psi^2,{\rm trunc}} \leq \sum^{\infty}_{j_1,\dots j_D = M} \left| \tilde{a}_{j_1\dots j_D} \right| \left| \frac{d T_{j_1} (x_1) }{d x_1}\right| \dots \left| \frac{d T_{j_D} (x_D) }{d x_D}\right| \leq \sum^{\infty}_{j_1,\dots j_D = M} \left| \tilde{a}_{j_1\dots j_D} \right| j_1^2  \dots  j_D^2
\end{align}
We have that
\begin{align}
    \sum^{\infty}_{j=M} r^{-j} j^2 = \mathcal{O} \left(r^{-M} M^2\right),
\end{align}
thus,
\begin{align}
    \epsilon_{\psi^2, {\rm trunc}} = \mathcal{O} \left( \max_{z\in D(\mathbf{r})}|\Psi(z)| M^{2D} r^{-M D}\right).
\end{align}
Now, we want the bound on $|\Psi(z)|$ on the polycylinder $D(\mathbf{r})$:
\begin{align}
    \max_{z\in D(\mathbf{r})}|\Psi(z)| = \max_{z\in D(\mathbf{r})}\left|  \int^{z_1}_{-1} \dots \int^{z_D}_{-1} \psi^2(z') d^D z' \right| \;\le\; (2r)^{D}\Lambda^2 \, e^{2 D\Lambda r}.
\end{align}
With this,
\begin{align}
    \epsilon_{\psi^2, {\rm trunc}} = \mathcal{O} \left( e^{2\log \Lambda}  e^{D \log{2r}}\, e^{2 D\Lambda r} e^{2D \log{M}} e^{-M D \log{r}}\right).
\end{align}
Solving for $M$, we get
\[
M
= \mathcal{O}\left(
-\frac{2}{\log r}\;
W\!\left(
-\frac{\log r}{2}\,
\exp\!\left(
\frac{\log \epsilon
      - 2\log \Lambda
      - D\log(2r)
      - 2D\Lambda r}{2D}
\right)
\right)
\right),
\]
where $W(x)$ represents the Lambert $W$ function.
Here, the relevant branch would be $W_{-1}(x)$ for $x\in [-\frac{1}{e},0]$, for which we have the following bound
\begin{align}
    -W_{-1}(x)  &< -\log(-x).
\end{align}

\begin{proof}
    We start off with the following inequality from \cite{Corless_LambertW_inequality}:
\begin{align}
        -W_{-1}(-e^{-u-1}) < 1 + \sqrt{2 u} + u
\end{align}
    for $u > 0$. A simpler but looser bound from this is
    \begin{align}
         -W_{-1}(-e^{-u-1}) < 1 + u.
    \end{align}
    We identify $\log(-x) = -u - 1$, thus the inequality in terms of $x$ is
     \begin{align}
         -W_{-1}(-e^{-u-1}) < -\log(-x)
    \end{align}
\end{proof}

Thus, using this bound we obtain:
\begin{align}
    M &= \mathcal{O} \left(-\frac{2}{\log{r}} \left({\log { \frac{\log{r}}{2}}} + \frac{\log \epsilon
      - 2\log \Lambda
      - D\log(2r)
      - 2D\Lambda r}{2D}\right)\right)
    \end{align}
Letting $r$ be $\mathcal{O}(1)$, and ignoring terms that grow like $\log{\Lambda}$, the expression simplifies to:
\begin{align}
      M&= \mathcal{O} \left( \frac{1}{D}\log\left(1/\epsilon\right) + \Lambda\right).
\end{align}
The error coming from the spilling of high-order contributions towards lower order coefficients (aliasing) in the interpolation procedure for $\Psi(x)$. That is, any higher-order contribution on the nodes gets projected onto the low-order coefficients. This, can be bounded through the inequality from \cite{Glau02112019}
\begin{align}
   \left| \Psi- I_M \right| = \epsilon_{\Psi,{\rm alias}} \leq   2\sum^{\infty}_{j_1,\dots j_D = M} \left| \tilde{a}_{j_1\dots j_D} \right|,
\end{align}
the truncation error
\begin{align}
        \left| \Psi - \Psi_M \right| = \epsilon_{\Psi,{\rm trunc}} \leq   \sum^{\infty}_{j_1,\dots j_D = M} \left| \tilde{a}_{j_1\dots j_D} \right|\cr
\end{align}
and the triangle inequality, obtaining:
\begin{align}
    \left| \Psi_M - I_M \right| \leq   3\sum^{\infty}_{j_1,\dots j_D = M} \left| \tilde{a}_{j_1\dots j_D} \right|.
\end{align}
Thus, because $T_{j}(1) =1$, the error on the determined coefficients (assuming perfect samples of $\Psi(x)$) is also bounded through
\begin{align}
    \sum^{M-1}_{j_1,\dots,j_D = 0} \left| \tilde{a} - \tilde{a}'\right| \leq 3\sum^{\infty}_{j_1,\dots j_D = M} \left| \tilde{a}_{j_1\dots j_D} \right|.
\end{align}
With this, the error from aliasing on the estimate of $\psi^2(x)$ is
\begin{align}
    \epsilon_{\psi^2,{\rm alias}}=\mathcal{O}\left( M^{2D} \epsilon_{\Psi,{\rm alias}}\right),
\end{align}
and we have that
\begin{align}
    \epsilon_{\Psi,{\rm alias}} \leq 3\sum^{\infty}_{j_1,\dots j_D = M} \left| \tilde{a}_{j_1\dots j_D} \right| = \mathcal{O} \left( \max_{z\in D(\mathbf{r})}|\Psi(z)|  r^{-M D}\right) = \mathcal{O}\left(\frac{\epsilon_{\psi^2,{\rm trunc}}}{M^{2D}}\right),
\end{align}
thus
\begin{align}
    \epsilon_{\psi^2,{\rm alias}} = \mathcal{O} \left(\epsilon_{\psi^2,{\rm trunc}}\right).
\end{align}

We can obtain the interpolation coefficients $a_{j_1 \dots j_D}$ (note normalization from \Cref{eq:interp}) by solving a system of $M^D$ linear equations, which takes $\mathcal{O}(M^{3D}) = \operatorname{polylog}(1/\epsilon)$ classical time.

The error propagated from the sample points at the moment of solving such a system is studied in many texts. This is related to the condition number of the linear system in question (See, for example, \cite{Rendon2024improvedaccuracy,Lebesgue_survey}). The error propagated, if the error on the samples is at most $\varepsilon$, would be $\mathcal{O}\left(\varepsilon \log^D (M)\right)$.

\begin{remark}
    While theoretically understanding how the errors propagate, fitting noisy data exactly might lead to unnecessary overfitting. Depending on the noise level, one might find it more accurate in practice to adopt some machine learning techniques, like using a least-squares fit with a lower degree Chebyshev polynomial, and adding an $\ell_1$ or $\ell_2$ norm penalization on the coefficients together with cross-validation (i.e.~using Lasso or Ridge regression).
\end{remark}

We will now explain how to extract the function and how the error on the integrals propagates towards the extracted function.

\subsection{Function extraction}

Now recall that we can differentiate the Chebyshev polynomials like \begin{align}
    \frac{\dd T_j(x)}{\dd x} = j \cdot U_{j-1}(x)\,,
\end{align} where $U_j(x)$ are the Chebyshev polynomials of the second kind. Hence also define \begin{align}
u_j(x) = \sqrt{2/M} \cdot U_j(x) 
\end{align} as the normalized versions of Chebyshev polynomials of the second kind, obeying \begin{align}
    \frac{\dd t_j(x)}{\dd x} = j \cdot u_{j-1}(x)\,,
\end{align}
for $j\geq 1$. We also have that \begin{align}
   \max_{x\in [-1,1]} |U_j (x)| \leq j+1,
\end{align}
thus, we know that
\begin{align}
    \max_{x\in[-1,1]}\left| \frac{{\rm d} t_j (x)}{{\rm d} x} \right| \leq \sqrt{2} j^{3/2} .
\end{align}
Having the interpolation \begin{align} \Psi(x_1,\dots,x_D) \approx\sum_{j_1,\dots, j_D = 0}^{M-1} a_{j_1\dots j_D} \cdot t_{j_1}(x_1) \dots t_{j_D}(x_D)\,,
\end{align} we may obtain an approximation to the encoded function $\psi$ from \begin{align}
    \psi(x_1,\dots,x_D) &= \left(\frac{\partial^D}{\partial x_1\dots \partial x_D} \Psi(x_1,\dots,x_D)\right)^{1/2}\\
    &\approx \left|\sum_{j_1,\dots, j_D = 1}^{M-1} a_{j_1\dots j_D} \cdot j_1 \dots j_D \cdot u_{j_1-1}(x_1) \dots u_{j_D-1}(x_D)\right|^{1/2}\,,\label{eqn:psi approximation}
\end{align} where we have additionally included the absolute value to ensure that this approximation is real.

The error incurred from the differentiation will yield an overhead scaling cubically with the number of summands $M^D$, $\epsilon_{\psi^2} = \mathcal{O}(M^{3D}\cdot \epsilon_\Psi)$.  To see this we look at:
Ignoring the truncation error from using a finite size $M$ interpolation, we know that the propagated uncertainty is upper bounded by:
\begin{align}
   \epsilon_{\psi^2} &= \left|\sum_{j_1,\dots, j_D = 1}^{M-1}(a-a^\star)_{j_1\dots j_D} \frac{\partial^D}{\partial x_1\dots \partial x_D} t_{j_1}(x_1) \dots t_{j_D}(x_D) \right| \cr 
   &\leq \sum_{j_1,\dots, j_D = 1}^{M-1}\left|(a-a^\star)_{j_1\dots j_D} \frac{\partial^D}{\partial x_1\dots \partial x_D} t_{j_1}(x_1) \dots t_{j_D}(x_D) \right| 
\end{align}
Now, using H\"older's inequality (or just a simple estimation lemma for sums) we obtain the following upper bound:
\begin{align}
    &\sum_{j_1,\dots, j_D = 1}^{M-1}\left|(a-a^\star)_{j_1\dots j_D} \frac{\partial^D}{\partial x_1\dots \partial x_D} t_{j_1}(x_1) \dots t_{j_D}(x_D) \right|  \cr 
    &\quad\leq \max \left| \frac{\partial^D}{\partial x_1\dots \partial x_D} t_{j_1}(x_1) \dots t_{j_D}(x_D)\right|\sum_{j_1,\dots, j_D = 1}^{M-1}\left|(a-a^\star)_{j_1\dots j_D} \right|  \cr
    & \quad\leq \left(\sqrt{2} M^{3/2}\right)^{D} \| (a- a^\star) \|_1 
\end{align}

We also have the following bound from the propagated error on the expansion coefficients from the error on the sampled function on the Chebyshev nodes:
\begin{align}
    \| a - a' \|_1 &= \| V_{\rm perturbed}^{-1} (f-f') \|_1\cr 
    & \leq\| V_{\rm perturbed}^{-1} \|_1 \| (f-f') \|_1 \cr
    &\leq M^{D}\| V_{\rm perturbed}^{-1} \|_1 \| (f-f') \|_{\infty} \cr 
    & \leq M^{3 D/2 } \| V_{\rm perturbed}^{-1} \|_2 \| (f-f') \|_{\infty} \cr
    &= \mathcal{O}\left( M^{3 D/2} \epsilon_{\Psi}\right).
\end{align}
Finally, we have
\begin{align}
    \epsilon_{\psi^2} = \mathcal{O}\left( 2^{D/2} M^{3D} \epsilon_{\Psi}\right).
\end{align}
Now, the error propagated from taking the square root will scale like $\epsilon_\psi = \mathcal{O}\left( \frac{\epsilon_{\psi^2}}{\min_x \psi(x)} \right)$. Since we can shift the function $\psi(x)$ so that $\min_x \psi(x)$ is bounded away from $0$, we can extract a representation of $\psi(x)$ within an error $\mathcal{O}(2 ^{D/2} M^{3D}\cdot \epsilon_\Psi)$. Hence by choosing $\epsilon_\text{cheb} \sim \epsilon_\text{QAE} \sim \epsilon_\text{total}$, we can get $\psi(x)$ within an error $\epsilon_\text{total}$ using $\tilde{O}\left(\frac{ 2^{D/2} M^{4D} }{\epsilon_{\rm total}} \right)$ queries of oracles $U_G$, $U_G^\dagger$, and $C\text{-}U_{G}$. Here $U_G$ is the Grover operator for estimating the integrated probabilities. The extra $M^D$ factor comes from the number of nodes required for the $D$-dimensional Chebyshev interpolation.

If we extend to the case where instead the circuit that prepares $\vert \psi\rangle$, does so with a sub-normalization:
\begin{align}\label{eq:subnorm_psi}
    \langle 0 |_a U_{\psi}|0 \rangle | 0 \rangle_a = a_{\psi} | \psi \rangle,
\end{align}
the cost scaling simply changes by a factor of $\frac{1}{a_{\psi}}$. We formalize our results in the following theorem:

\begin{theorem}\label{thm:SE_multidim}
Provided an unknown analytic function function $\psi(x)$ for $x \in [-1,1] $ that is normalized:
\begin{align*}
\int_{-1}^{1} \dots \int_{-1}^{1} |\psi(y_1,\dots,y_D)|^2\ \dd y^D\ =1,.
\end{align*}
and whose derivatives are bounded by $\|\partial^\alpha \psi \|_\infty \leq \Lambda^{|\alpha|+1}$, which can be stored in a quantum memory with $n$ qubits the following way:
\begin{align*}
    \langle 0 |_a U_{\psi}|0 \rangle | 0 \rangle_a = a_{\psi} | \psi \rangle,
\end{align*}
where
\begin{align*}
    |\psi\rangle = \frac{1}{\sqrt{\mathcal{N}}} \sum_{j_1,\dots,j_D = 0}^{2^n-1} \psi\left(x^{(j_1)},\dots,x^{(j_D)}\right) |j_1\rangle \dots |j_D\rangle\,,
\end{align*}
 $\mathcal{N} = \sum_{j_1,\dots,j_D = 0}^{2^n-1}|\psi\left(x^{(j_1)},\dots,x^{(j_D)}\right) |^2 $, and $x^{(j)} = 2 j/2^n -1$, one can estimate it at a quantum gate cost that scales like:
 \begin{align*}
    \tilde{O}\left(\frac{1}{a_{\psi}}\frac{ 2^{D/2} M^{4D} }{\epsilon_{\rm total}} \right).
\end{align*}
where $M= \mathcal{O}\left(\Lambda + \frac{1}{D}\log{1/\epsilon_{\rm total}}\right)$.
\end{theorem}

%%%%%%%%%%%%%%%%%%%%%%%%%%%%%%%%%%%%%%%%%%%%%%%%%%%%%%%%%%%%%%%%%%%%%%%%%%%%%%%%

\section{Numerical simulations}

In this section, we trial the performance of our approach numerically by carrying out a state vector simulation of the quantum circuits using \textsc{pytket}. To do that, we firstly encode a given function $\psi$ into the amplitudes of a quantum state using the Grover-Rudolph method \cite{grover2002} (note that this step is not efficient for general functions; however, here we are rather concerned with the subsequent extraction). Then we proceed to draw samples of the square integral $\Psi$ using our method, which we fit with Chebyshev polynomials, and finally extract an estimate to $\psi$. Due to the restrictions of a classical simulation, we will consider only the one dimensional case.

\subsection{Grover-Rudolph technique for function encoding}

We will consider the example $\psi(x) \propto (\sin(5x)+2)\cdot e^{x}$ up to a normalization factor ensuring $\int_{-1}^1\psi(x)^2\ \dd x = 1$. To construct the unitary $U_\psi$ for preparing $\ket{\psi}$ we will be using the Grover-Rudolph method, where the oracle for calculating the necessary rotation angles would be synthesized from their values obtained by classical numerical integration. The final precision of the full encoding would hence depend on the number of qubits $n$ storing the amplitudes, which controls the sampling frequency of the function, as well as the number of ancilla qubits $m$ used for storing the rotation angles and hence controlling the accuracy of the sampling.

\begin{figure}[H]
  \centering
  \begin{subfigure}{0.48\linewidth}
    \centering
    \includegraphics[width=\linewidth]{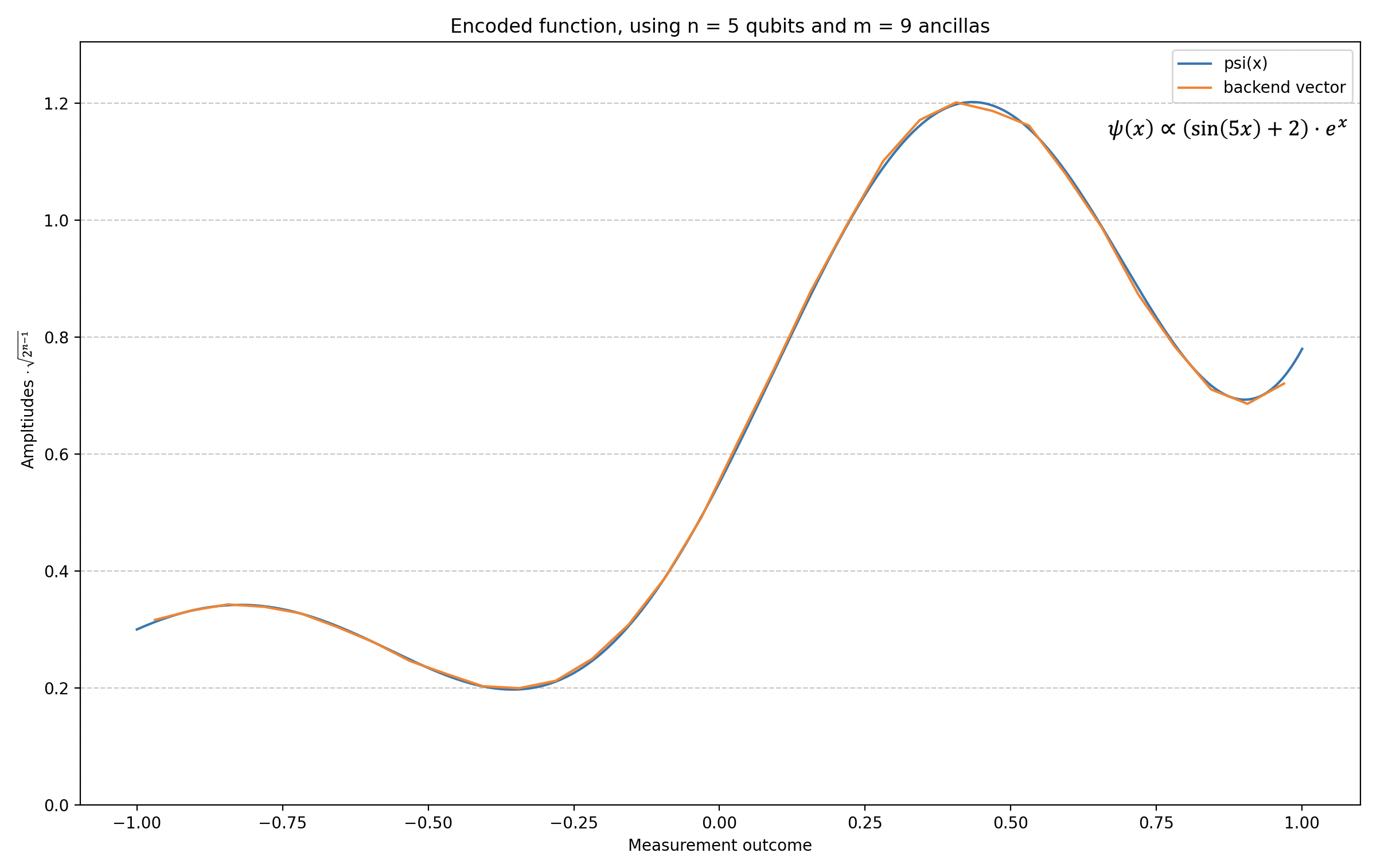}
    \caption{using $m=9$ ancillas for precision}
  \end{subfigure}\hfill
  \begin{subfigure}{0.48\linewidth}
    \centering
    \includegraphics[width=\linewidth]{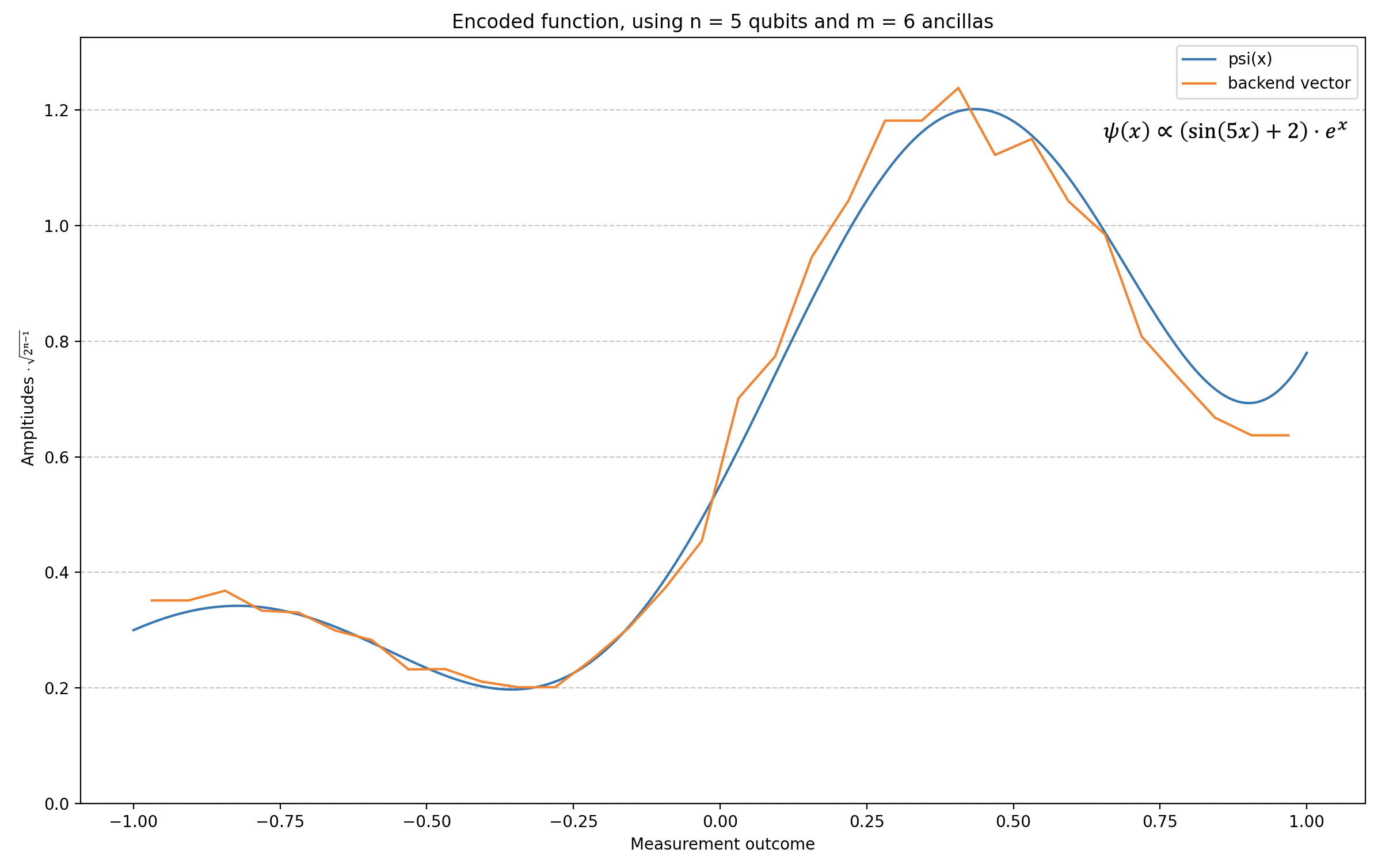}
    \caption{using $m=6$ ancillas for precision}
  \end{subfigure}
  \caption{Encoding the function $\psi(x)\propto (\sin(5x)+2)e^x$ into a quantum state with $n=5$ qubits using the Grover–Rudolph method with different numbers of ancilla qubits controlling the precision, and comparing the rescaled amplitudes with the exact functional values.}\label{fig:Grover-Rudolph}
\end{figure}

On Figure \ref{fig:Grover-Rudolph}, we compare the exact function to the amplitudes of the encoded state (rescaled by $\sqrt{2^{n-1}}$ to match the scale), and consider the effects of different numbers $m$ of ancilla qubits controlling the sampling precision of the encoded function. In both cases we store the functional values in $n = 5$ qubits. We see that $m = 9$ ancilla qubits already give us a faithful representation of the exact function, however, because of the classical complexity of the subsequent simulation, we will have to restrict ourselves to the less accurate encoding using $m = 6$ ancilla qubits when sampling the cumulative distribution in Section \ref{sec:numerics_integral}.

\subsection{Integral estimation}\label{sec:numerics_integral}

Given the oracle $U_\psi$ for encoding the function $\psi$ into the state $\ket{\psi}$, we may implement the necessary Grover operator \eqref{eqn:Grover operator} for the amplitude estimation. To do that, we also require the quantum comparator, which shall be constructed by ripple-carry addition \cite{cuccaro2004newquantumripplecarryaddition}. We shall use another $K = 5$ qubits for the precision of quantum phase estimation, together with a simple majority vote from many shots.

\begin{figure}[H]
    \centering
    \includegraphics[width=\linewidth]{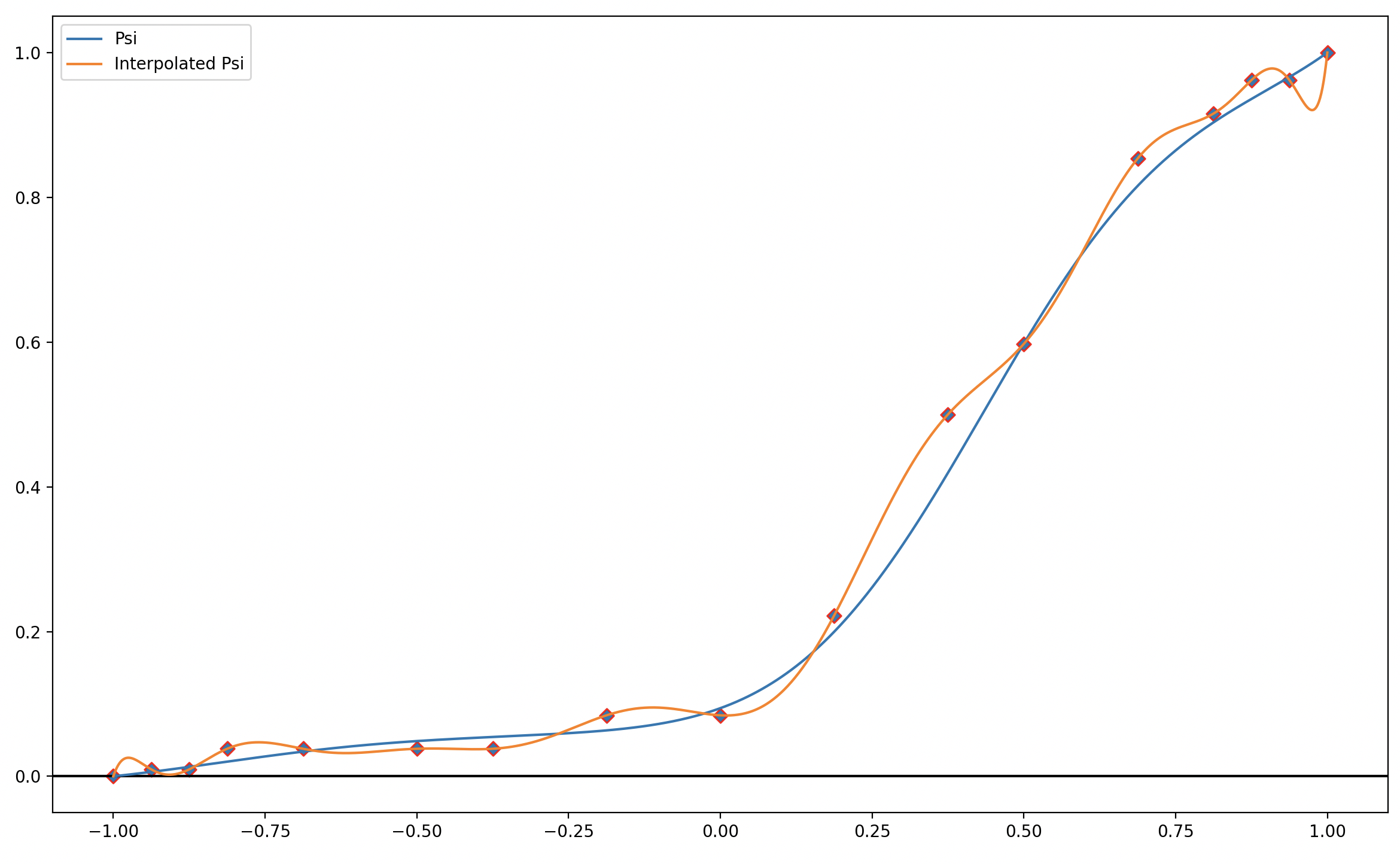}
    \caption{Measuring and interpolating the integral $\Psi(x) \propto \int_{-1}^x (\sin(5y)+2)e^y\ \dd y$ with the described technique, using the encoded function $\psi(x) \propto (\sin(5x)+2)e^x$ in $n=5$ qubits with $m=6$ ancillas, and QPE with $K=5$ qubits for precision, and $M=17$ Chebyshev nodes (or their closest $n$-bit approximations). The figure shows individual measured points, and compares the function interpolated from them with the exact integral $\Psi(x)$.}
    \label{fig:Psi_integral_measured}
\end{figure}

On Figure \ref{fig:Psi_integral_measured}, we compare the exact values of $\Psi(x)$ to that of $M = 17$ points sampled using our method from the previously encoded state $\ket{\psi}$. The measured values match the exact ones up to a reasonable amount of error introduced by the inaccurate encoding and quantum amplitude estimation. Further, we fit the $M$ sampled values with Chebyshev polynomials (exactly) and compare this interpolated integral to $\Psi$. Note that, while $\Psi$ has to be monotonically increasing because of the positiveness of $\psi$, the interpolated integral need not obey this restriction. One might consider different interpolation techniques, including cross-validation and imposing the monotonicity condition, to potentially see a better performance in practice.

\subsection{Function extraction}

\begin{figure}[H]
    \centering
    \includegraphics[width=\linewidth]{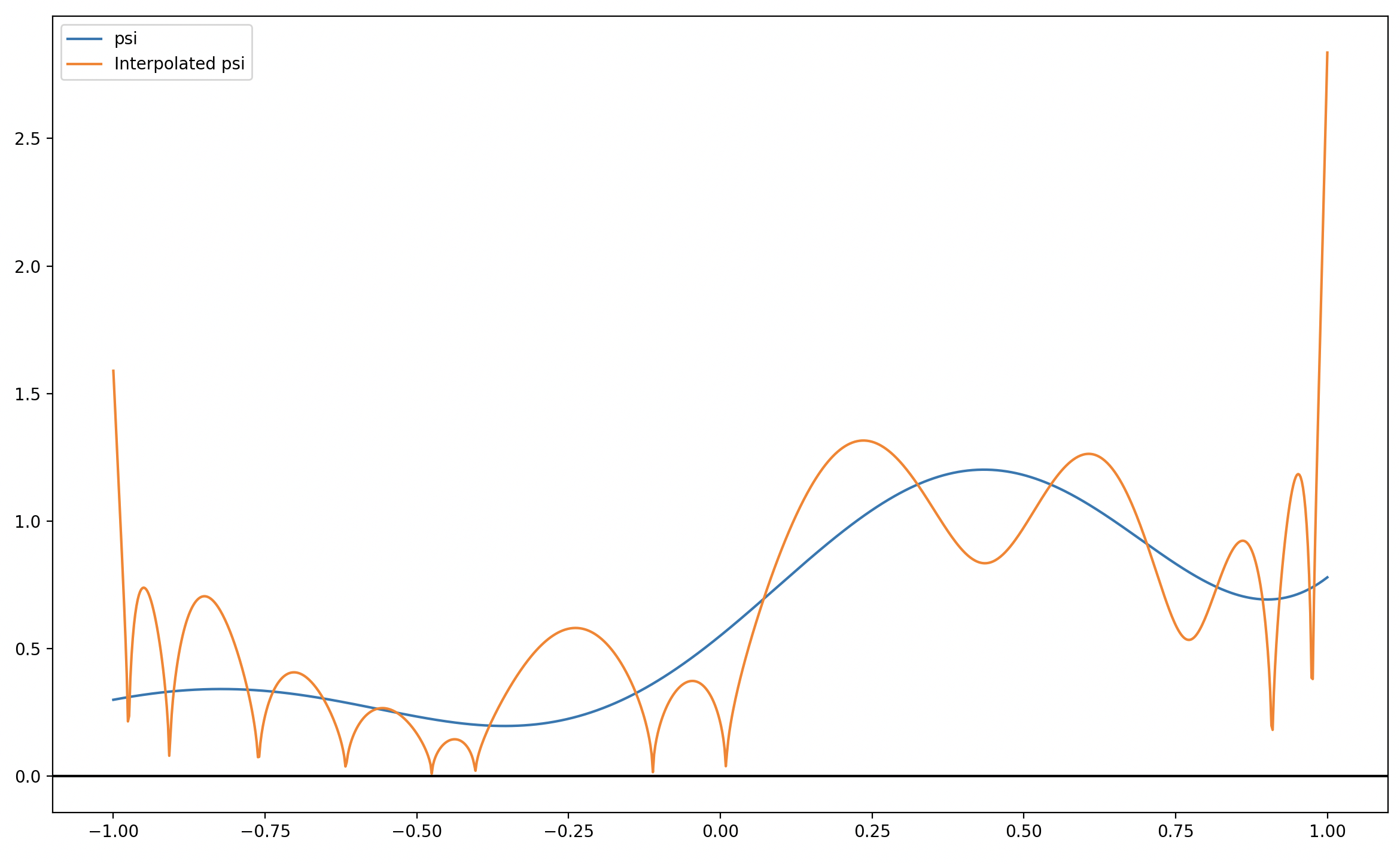}
    \caption{Function extracted from the measured integral values on Figure \ref{fig:Psi_integral_measured}. This step incurs errors proportional to the number of points $M$ cubed. The precision in the measured values of $\Psi(x)$ in this small scale simulation wasn't sufficient in order to yield a reasonable estimate for $\psi(x)$.}
    \label{fig:extracted_function}
\end{figure}

After obtaining the Chebyshev coefficients from interpolating $\Psi$, we simply extract a representation of the function $\psi$ using Equation \eqref{eqn:psi approximation}. We compare this extracted function with the exact values of $\psi$ on Figure \ref{fig:extracted_function}. Here we can see that the error incurred in this last step, which scales with $M^3$, is simply too large to obtain a reasonable representation of $\psi$. Reducing this error by orders of magnitude would require only a several more qubits, which, however, is outside of the scope of this small scale simulation.

\subsection{Illustrative function extraction with controlled errors}

\begin{figure}[H]
    \centering
    \subfloat[integral $\Psi(x)$, $100\times$ smaller errors]{
            \includegraphics[width=0.5\textwidth]{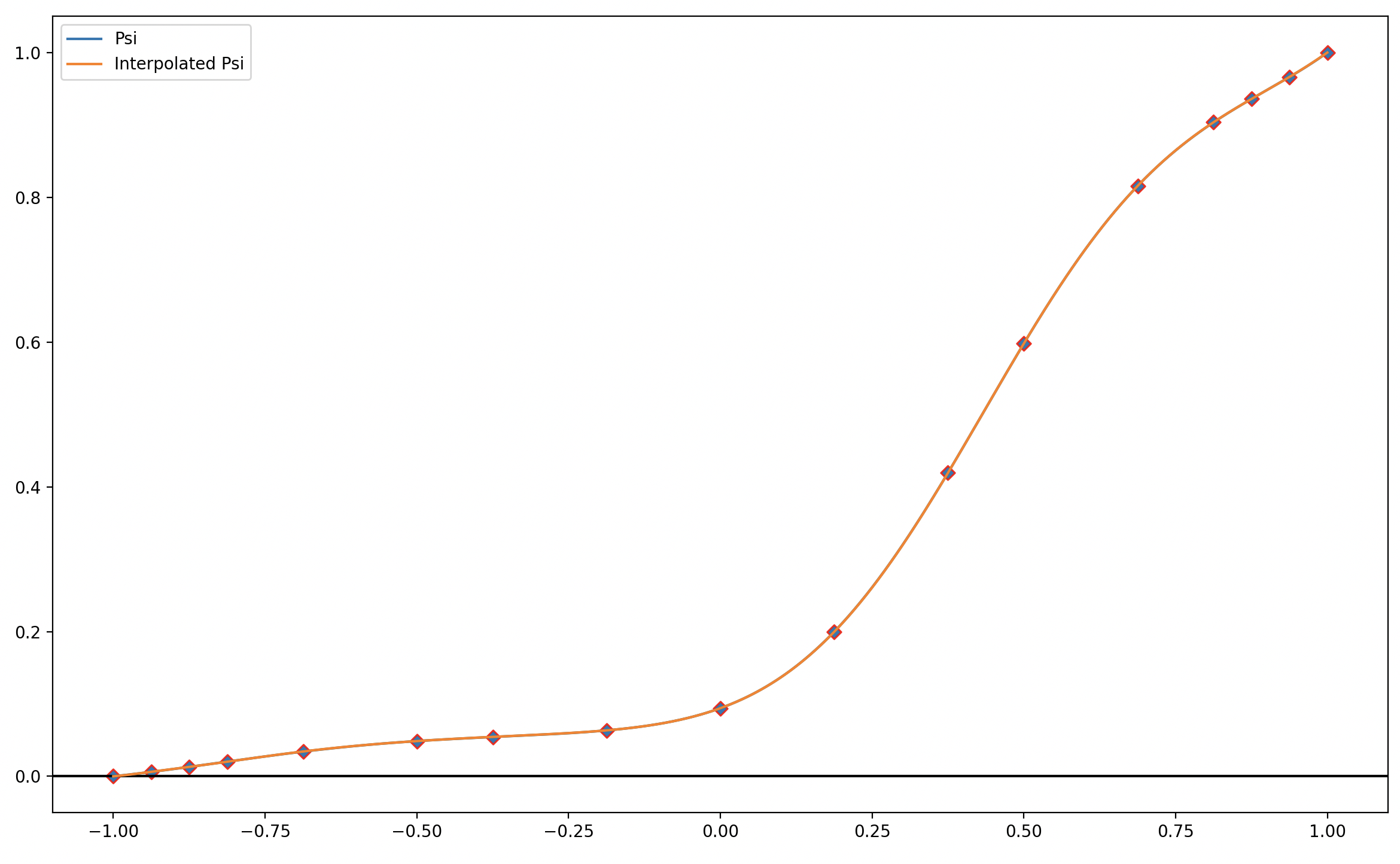}}
    \subfloat[function $\psi(x)$, $100\times$ smaller errors]{
            \includegraphics[width=0.5\textwidth]{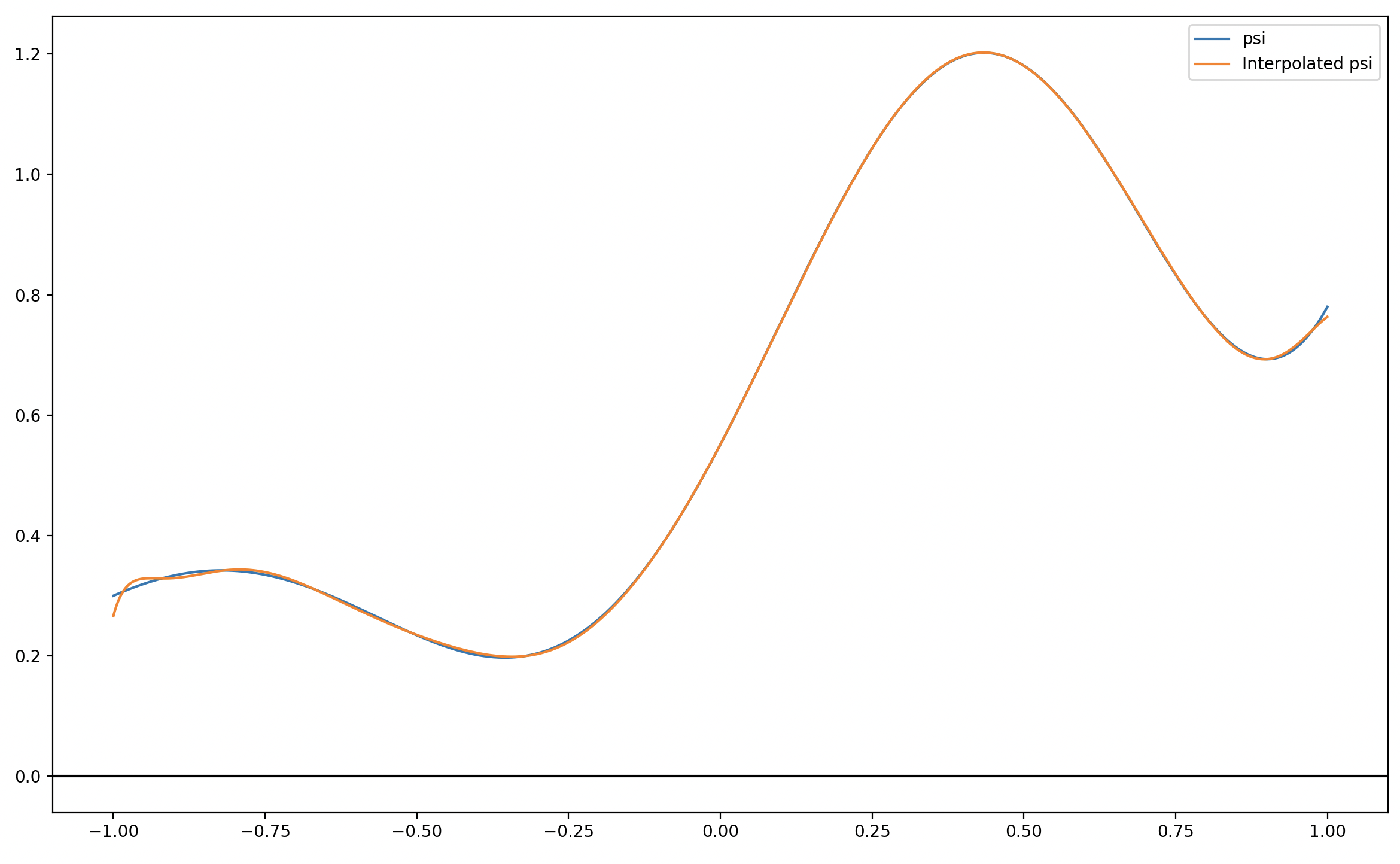}}\\
    \subfloat[integral $\Psi(x)$, $20\times$ smaller errors]{
            \includegraphics[width=0.5\textwidth]{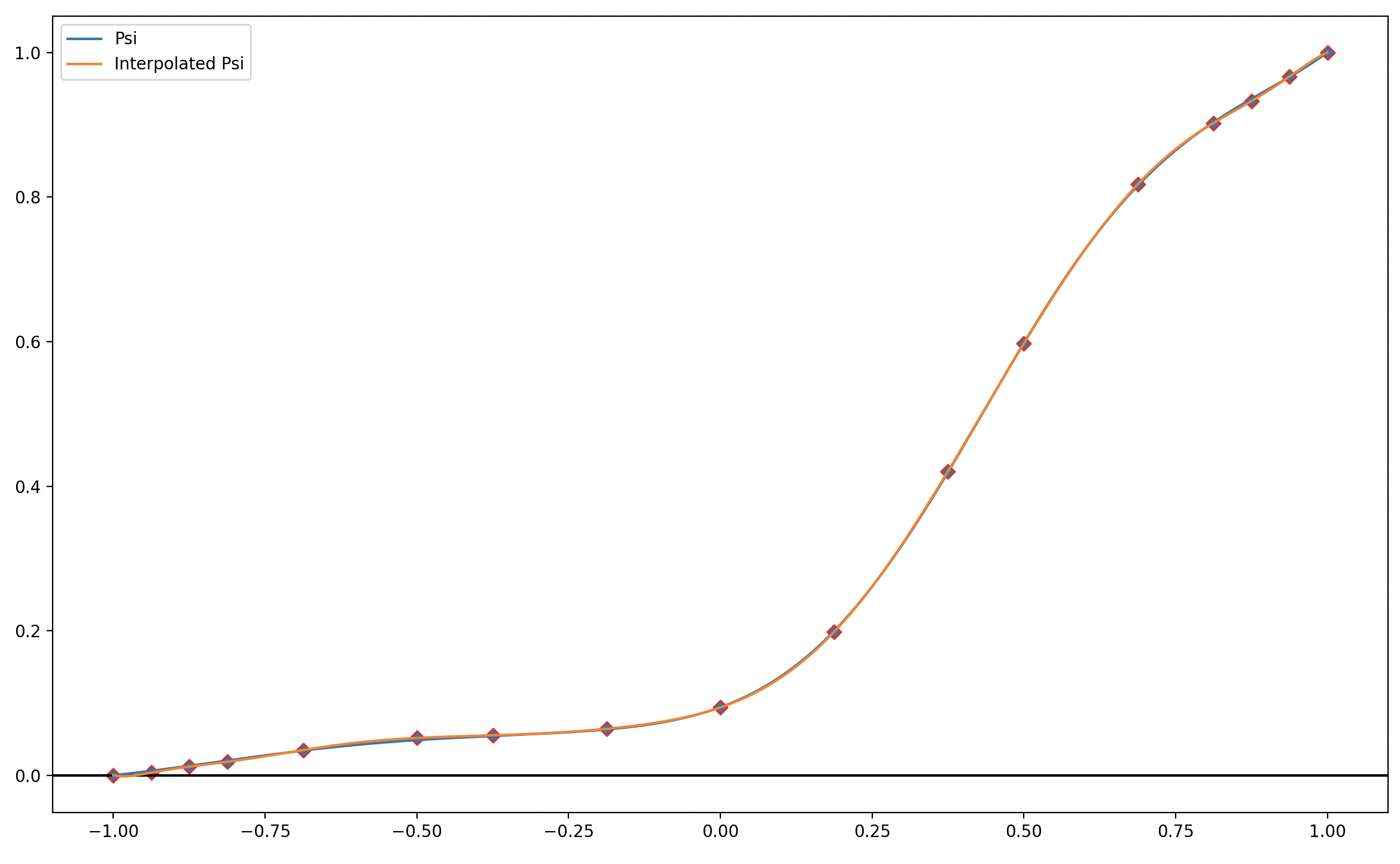}}
    \subfloat[function $\psi(x)$, $20\times$ smaller errors]{
            \includegraphics[width=0.5\textwidth]{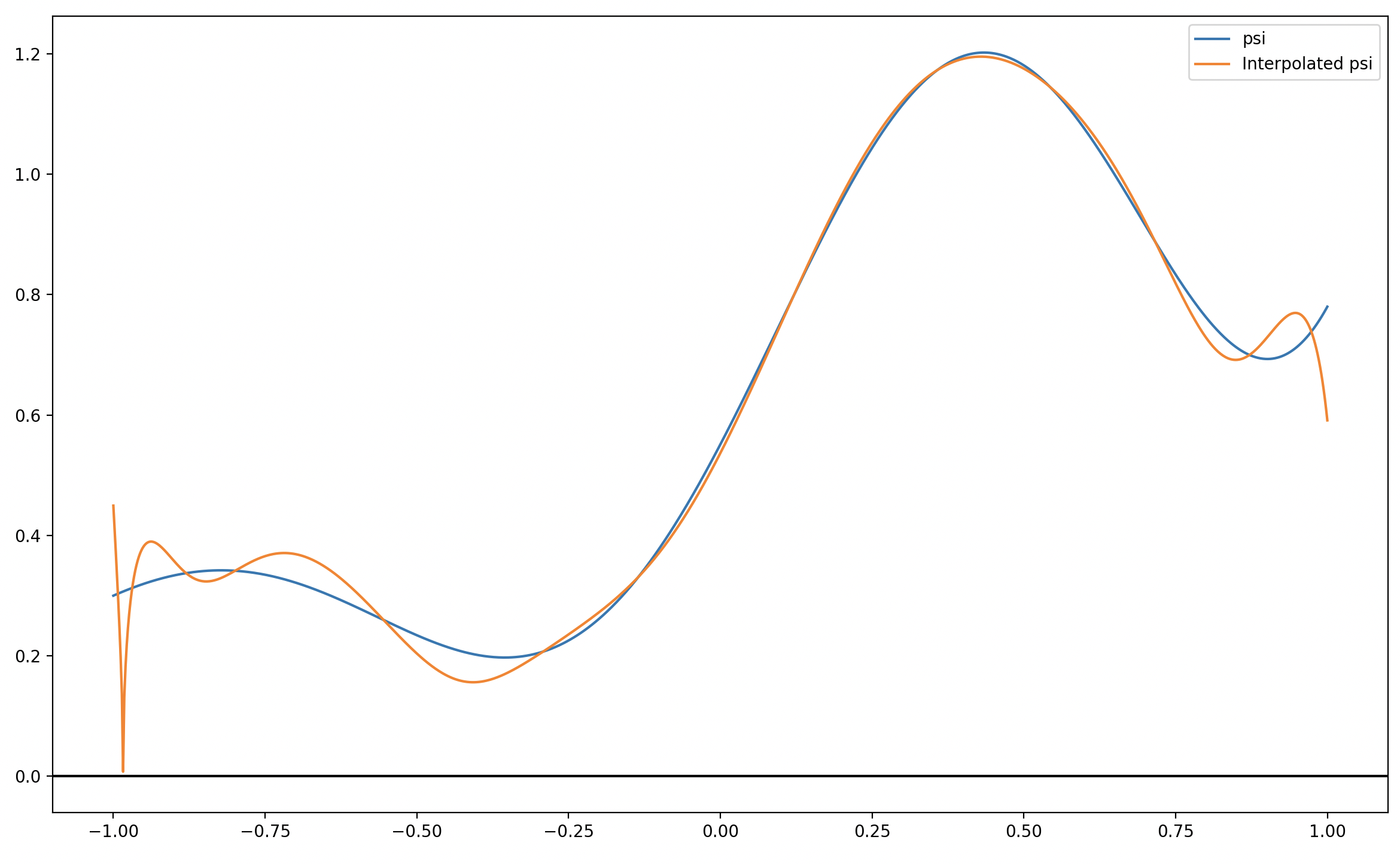}}
    \caption{Demonstration of extracting the function $\psi(x)$ from the values of $\Psi(x)$ sampled at the same $M=17$ Chebyshev nodes as in Figure \ref{fig:Psi_integral_measured}, but instead of a quantum circuit simulation, here we sample $\Psi(x)$ exactly and add random Gaussian noise to it, which we control to be roughly $100\times$ or $20\times$ smaller than the one we measured respectively. These precisions would require only a several more qubits in each register, which we, however, can't simulate.}
    \label{fig:fake_extraction}
\end{figure}

On Figure \ref{fig:fake_extraction}, to circumvent the limitations of the classical simulations in order to illustrate the solution extraction with higher accuracy, we simply sample the values of $\Psi$ exactly at the same $M = 17$ Chebyshev nodes, and modify them with random Gaussian noise in a controlled manner. After that, we extract the representation of $\psi$ as before. Here we can see that reducing the noise on $\Psi$ by a factor of $20$ already gives a reasonable approximation on $\psi$ throughout most of the domain, while reducing it by a factor of $100$ gives a highly accurate representation of $\psi$.

%%%%%%%%%%%%%%%%%%%%%%%%%%%%%%%%%%%%%%%%%%%%%%%%%%%%%%%%%%%%%%%%%%%%%%%%%%%%%%%%

\section{Conclusions and outlook}

In this work we have extended the results of \cite{rendon2025_SE}, which paves the way for fully solving differential equations and linear systems on quantum computers with the mild assumptions of the solution having a smoothness parameter $\Lambda$ in the characterization with the bound: $\|\partial^\alpha \psi \|_\infty \leq \Lambda^{|\alpha|+1}$. This condition encompasses a large class of problems in finance, fluid dynamics, heat dynamics, to name a few. We hope that this leads to more instances of quantum advantage in this kind of problems given that we improve an important bottle neck with guarantees of extracting the solution within a target error. The mild dimensionality curse in the number of variables is expected for any method relying on multi-variable interpolation methods even in the context of classical methods. However, our method still holds an exponential improvement in cost with respect to the number of qubits encoding the solution over the naive way of applying Quantum Amplitude Estimation. Moreover, because it relies on a linear regression for the interpolation of the accumulated probabilities, the cost bounds we provide are guaranteed.
We hope to improve this in the future through sparser grids for interpolation or possibly improve heuristically through neural/tensor networks although guarantees of convergence  to an optimum are limited for this last kind of method.
Also, an efficient way to handle shocks, kinks and other types of singularities would enlarge the types of solutions one can efficiently extract from a quantum register.

%%%%%%%%%%%%%%%%%%%%%%%%%%%%%%%%%%%%%%%%%%%%%%%%%%%%%%%%%%%%%%%%%%%%%%%%%%%%%%%%
%\nocite{*}

\bibliographystyle{naturemag}

%%%%%%%%%%%%%%%%%%%%%%%%%%%%%%%%%%%%%%%%%%%%%%%%%%%%%%%%%%%%%%%%%%%%%%%%%%%%%%%%

\end{document}